\documentclass[useAMS,usenatbib,referee]{mn2e}
\usepackage[koi8-r]{inputenc}
\usepackage[english]{babel}
\usepackage{graphicx}
\usepackage{amssymb, amsmath}
\usepackage{epsf}

\def\la{\; \raise0.3ex\hbox{$<$\kern-0.75em\raise-1.1ex\hbox{$\sim$}}\;}
\def\ga{\;  \raise0.3ex\hbox{$>$\kern-0.75em\raise-1.1ex\hbox{$\sim$}}\;}

\def\pFn{p_{\raise-0.3ex\hbox{{\scriptsize F$\!$\raise-0.03ex\hbox{\rm n}}}}
}  
\def\pFp{p_{\raise-0.3ex\hbox{{\scriptsize F$\!$\raise-0.03ex\hbox{\rm p}}}}
}  
\def\pFe{p_{\raise-0.3ex\hbox{{\scriptsize F$\!$\raise-0.03ex\hbox{\rm e}}}}
}  
\def\pFmu{p_{\raise-0.3ex\hbox{{\scriptsize F$\!$\raise-0.03ex\hbox{
\rm $\mu$}}}} }  
\def\m@th{\mathsurround=0pt }
\def\eqalign#1{\null\,\vcenter{\openup1\jot \m@th
   \ialign{\strut$\displaystyle{##}$&$\displaystyle{{}##}$
   \hfil \crcr#1\crcr}}\,}
\title[The neutrino emission due to plasmon decay
and neutrino luminosity of white dwarfs]
{The neutrino emission due to plasmon decay
and neutrino luminosity of white dwarfs}
\author[E. M. Kantor, M. E. Gusakov]
{E. M. Kantor$^{}$\thanks{E-mail:
kantor@mail.ioffe.ru (EMK); gusakov@astro.ioffe.ru (MEG)},
 M. E. Gusakov$^{}$\\
$^{}$Ioffe Physical Technical Institute,
Politekhnicheskaya 26, St.-Petersburg 194021, Russia}
\begin{document}

\date{Accepted 2007 August 14. Received 2007 August 10;
in original form 2007 July 04}

\pagerange{\pageref{firstpage}--\pageref{lastpage}} \pubyear{2007}

\maketitle

\label{firstpage}

\begin{abstract}
One of the effective mechanisms of neutrino energy losses
in red giants, presupernovae and in the cores of white dwarfs
is the emission of neutrino-antineutrino pairs
in the process of plasmon decay.
In this paper, we numerically calculate
the emissivity due to plasmon decay in a wide range
of temperatures $(10^7-10^{11})$ K and densities
$(2 \times 10^2-10^{14})$ g cm$^{-3}$.
Numerical results
are approximated by convenient
analytical expressions.
We also calculate and approximate
by analytical expressions the neutrino luminosity
of white dwarfs due to plasmon decay,
as a function of their mass
and internal temperature.
This neutrino luminosity depends
on the chemical composition of white dwarfs only
through the parameter $\mu_{\rm e}$
(the net number of baryons per electron)
and is the dominant neutrino luminosity in all white dwarfs
at the neutrino cooling stage.
\end{abstract}


\section{Introduction}

It is well known that neutrino emission plays
an important role in the evolution
of red giants, presupernovae,
white dwarfs, and neutron stars.
Neutrinos appear in a number of reactions
in dense stellar matter
(see, e.g., Yakovlev et al.\ 2001)
and freely escape from the star,
producing a powerful mechanism of their cooling.
One of the effective neutrino generation
mechanisms is the plasmon decay.

In contrast to ordinary photons in vacuum,
plasmons, which are quanta of electromagnetic field in a plasma,
can be not only transverse
(in this case two polarization vectors of plasmon
are perpendicular to wave vector),
but longitudinal as well.
The longitudinal plasmons
appear in the theory as a result of
quantization of the well known
Langmuir plasma waves.

Plasmon can decay into a neutrino-antineutrino pair,
$\gamma \rightarrow \nu + \overline{\nu}$.
The appropriate neutrino emissivity
was analyzed in a series of papers since 1963,
when Adams, Ruderman, and Woo
had suggested this mechanism of energy losses in dense
stellar matter for the first time.
An account of these papers and references
can be found in the review by Yakovlev et al.\ (2001)
as well as in a recent paper by Odrzywo\l{}ek (2007).
Here we discuss in more detail only three papers which
summarize and extend the results of previous works.

Itoh et al.\ (1992) calculated the emissivity due to plasmon decay
as a function of temperature and density and presented a table
of numerical values and an approximate fit formula.
Unfortunately, this approximate formula does not reproduce
analytical asymptotes for the emissivity and thus can be applied
only in a restricted region of temperatures and densities
(near the maximum value of the emissivity).
In addition, when calculating the emissivity,
Itoh et al.\ (1992) used approximate expressions
for the dielectric functions of electron gas
and for plasmon dispersion relations
which can be justified only
at low enough temperatures
(in a strongly degenerate electron gas).

On the contrary, Braaten $\&$  Segel (1993)
started with the most general expressions
for the neutrino emissivity
due to plasmon decay.
They did not make any assumptions concerning
degeneracy of the electron gas at calculating
the dielectric functions and plasmon dispersion relations.
To simplify their analysis, Braaten $\&$ Segel (1993)
suggested an elegant scheme to calculate approximately
the dielectric functions,
dispersion relations, and the neutrino emissivity.
However, these authors did not present any tables with
their numerical results or
any approximate formula for the emissivity.
Therefore, it is difficult to use their results
in applications.

Using the approximate method of Braaten $\&$ Segel,
Haft et al.\ (1994) calculated the emissivity due to plasmon decay
and fitted it by an analytical formula.
This formula accurately describes the emissivity in
a range
of temperatures and densities where the plasmon decay
is the most important neutrino emission mechanism.
However, the fitting expression of Haft et al.\ (1994)
does not satisfy the analytical asymptotes for the emissivity
(they are presented in Section~2).

In this paper we would like to fill in the gaps
in the literature devoted to the subject.
We will
({\it i}) numerically calculate
the neutrino emissivity due to plasmon decay
making no assumptions concerning
degeneracy or relativity of the electron gas;
({\it ii}) employ the approximate scheme of Braaten $\&$ Segel
and find a fitting expression for the emissivity
which reproduces the correct asymptotes.
Thus, our main goal is to facilitate the use of the data
on the neutrino emission
due to plasmon decay.

The paper is organized as follows.
In Section 2 we present general equations
describing the neutrino energy loss rate owing to plasmon decay.
In Section 3 we give the fit expression for
the plasma frequency which is a key parameter because
the asymptotes of the emissivity depend on it.
In Section 4 we present
the fit expressions for the emissivity.
In Section 5 we apply the results of the preceding sections
and find an analytical formula describing the neutrino
luminosity of white dwarfs as a function of their mass
and internal temperature.
We summarize in Section 6.
In Appendix A we present expressions
for the dielectric functions of the
electron-positron plasma.
Finally, in Appendix B we describe a table of
our numerical results.

\section{General equations}


The neutrino emissivity due to plasmon decay
can be presented as a sum of three components:
the longitudinal component $Q_{\rm l}$
(due to decay of longitudinal plasmons);
the transverse component $Q_{\rm t}$
(the decay of transverse plasmons governed by
the vector part of the weak interaction Hamiltonian);
and the axial component $Q_{\rm A}$
(the decay of transverse plasmons
governed by the axial part
of the weak interaction Hamiltonian).
The component $Q_{\rm A}$ is small and can be neglected
(see, e.g., Kohyama et al.\ 1994).

The emissivities $Q_{\rm t}$ and $Q_{\rm l}$ (per unit volume)
are given in the form of integrals
(see, e.g., Braaten 1991, Braaten $\&$ Segel~1993)
\begin{eqnarray}
Q_{\rm t} &=& 2 \, Q_{\rm 0} \, {\hbar^9 \over m_{\rm e}^9 c^{15}} \,
\int_0^\infty  dk \, k^2 Z_{\rm t}(k)
\left( \omega_{\rm t}^2-k^2 c^2 \right)^3 \, n_{\rm B}(\omega_{\rm t}),
\label{QT} \\
Q_{\rm l} &=& Q_{\rm 0} \, {\hbar^9 \over m_{\rm e}^9 c^{15}} \,
\int_0^{k_{\rm max}} dk \,
k^2 Z_{\rm l}(k) \left( \omega_{\rm l}^2-k^2 c^2 \right)^3 \,
n_{\rm B}(\omega_{\rm l}).
\label{QL}
\end{eqnarray}
Here, the integration is carried over
the plasmon wave number $k$.
In equations (\ref{QT}) and (\ref{QL})
$Q_{\rm 0} =\left[(m_{\rm e} c)^9/\hbar^{10} \right] \,\,
\left[G_{\rm F}^2/(96 \pi^4 \alpha) \right] \,\,
(\sum_\nu C_{\rm V}^2) \approx
1.3858 \times 10^{21}$ ${\rm erg \, s^{-1} \, cm^{-3}}$;
$G_{\rm F}=1.436 \times 10^{-49}$ erg cm$^3$
is the Fermi weak coupling constant;
$\alpha=e^2/(\hbar c) \approx 1/137$ is the fine structure constant;
$e$ and $m_{\rm e}$ are the electron charge and mass, respectively;
$\hbar$ is the Planck constant;
$c$ is the speed of light;
$\sum_\nu C_{\rm V}^2 \approx 0.9248$
is the sum of squared normalized
vector constants $C_{\rm V}$ over all neutrino flavors.
Furthermore, $\omega_{\rm t}(k)$ and $\omega_{\rm l}(k)$
are, respectively, the frequencies
of transverse and longitudinal plasmons,
which depend on the wave number $k$;
$Z_{\rm t}(k)^{-1} \equiv \partial(\omega_{\rm t}^2
\epsilon_{\rm t})/\partial(\omega_{\rm t}^2)$;
$Z_{\rm l}(k)^{-1} \equiv (\omega_{\rm l}^2 - k^2 c^2) \,
\partial \epsilon_{\rm l}/\partial (\omega_{\rm l}^2$),
where $\epsilon_{\rm t}$ and $\epsilon_{\rm l}$
are the transverse and longitudinal dielectric functions
of the electron-positron plasma, respectively.
Finally, $n_{\rm B}(\omega_{\rm t,l})
=1/\{ {\rm exp}[\hbar \omega_{\rm t,l}/(k_{\rm B} T)]-1\}$
is the Bose-Einstein distribution function
for transverse or longitudinal plasmons;
$T$ is the temperature; $k_{\rm B}$ is the Boltzmann constant;
$k_{\rm max}$ is the maximum wave number at which the
decay of longitudinal plasmon is still kinematically allowed
by energy and momentum conservation laws.

In the astrophysical literature
the emissivity is presented
as a function of temperature $T$
and the effective mass density $\tilde{\rho}$, given by
\begin{equation}
\tilde{\rho} \equiv
\rho/\mu_{\rm e},
\label{rho}
\end{equation}
where $\rho$ is the actual mass density;
$\mu_{\rm e}=\sum_i A_i n_i/(\sum_i Z_i n_i)$
is the net number of baryons per electron;
$Z_i$ and $A_i$ are, respectively,
the charge and mass numbers
of atomic nucleus species $i$;
$n_i$ is the number density
of these species.
Notice, that
at densities higher than
the neutron drip density
$\rho_{\rm d} \approx 4 \times 10^{11}$ g cm$^{-3}$,
free neutrons must be taken into account in the sum over $i$,
in addition to atomic nuclei,
when calculating $\mu_{\rm e}$.

It is straightforward to verify that
$\tilde{\rho}$ can be rewritten as
\begin{equation}
\tilde{\rho} \approx (n_{\rm e} -n_{\rm e^+}) \, m_{\rm u}.
\label{ne}
\end{equation}
Here, $n_{\rm e}$ and $n_{\rm e^+}$ are
the number densities of electrons and positrons;
$m_{\rm u}$ is the atomic mass unit.

The dependence of the emissivity $Q=Q_{\rm t} + Q_{\rm l}$
on $\tilde{\rho}$ for temperatures
$T=10^7, 10^8, 10^9$, and $10^{10}$ K
is presented in Fig.\ 1.
As seen from the figure,
at fixed $\tilde{\rho}$
the emissivity increases
with the growth of $T$.
If we fix $T$,
the dependence $Q(\tilde{\rho})$ has a maximum.
In the vicinity of the maximum
the plasma frequency of the electron-positron plasma
$\omega_{\rm p}$ is of the order of temperature,
$\hbar \omega_{\rm p} \sim k_{\rm B} T$
(see Section 5 for details).
At high temperatures and low densities
the emissivity ceases to depend on $\tilde{\rho}$
(see equations \ref{Wasy2}, \ref{QTasy1}, and \ref{QLasy1} below).
In the figure this situation
is illustrated by the upper curve,
which is plotted for $T=10^{10}$ K.
One sees that at ${\tilde \rho} < 10^8$ g cm$^{-3}$
the curve tends to be horizontal.

\begin{figure}
\begin{center}
\leavevmode
\epsfxsize=8cm
\epsfbox[60 200 560 670]{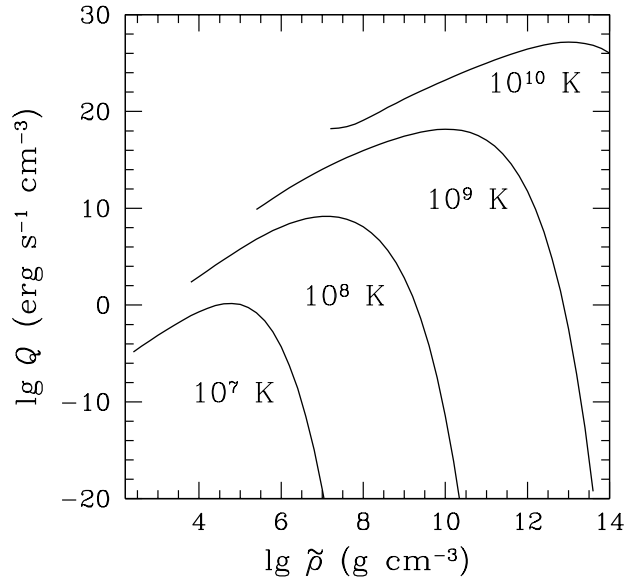}
\end{center}
\caption{
The emissivity $Q=Q_{\rm t}+Q_{\rm l}$
versus $\tilde{\rho}$ for
$T=10^7, 10^8, 10^9$, and $10^{10}$ K.}
\label{fig:fig1}
\end{figure}
As follows from equations (\ref{QT}) and (\ref{QL}),
for calculation of $Q_{\rm t}$ and $Q_{\rm l}$
one needs to know the dispersion relations for transverse and longitudinal
plasmons, $\omega_{\rm l}(k)$ and $\omega_{\rm t}(k)$,
as well as the dielectric functions
$\epsilon_{\rm t}(\omega,k)$ and $\epsilon_{\rm l}(\omega,k)$.
We calculated the dielectric functions
$\epsilon_{\rm t}(\omega, k)$ and $\epsilon_{\rm l}(\omega, k)$
for a wide range of densities and temperatures
in the random phase approximation
and numerically obtained the dispersion relations
and the plasma frequency $\omega_{\rm p}$.
The equations we used to compute the dielectric functions
of the electron-positron plasma are given in Appendix A.
These results were applied to calculate the integrals
(\ref{QT}) and (\ref{QL}).
In these calculations,
we did not make any simplifying assumptions
concerning the degree of degeneracy
or relativity of the electron gas.
The table with our numerical results can be found
on the web,
http://www.ioffe.ru/astro/NSG/plasmon/table.dat
(file table.dat).
This table is described in Appendix B.

The emissivities $Q_{\rm l}$ and $Q_{\rm t}$
depend on two parameters
characterizing stellar matter.
For example, one may choose
$T$ and $n_{\rm e}$
or $T$ and ${\tilde \rho}$
as proper parameters.
Following previous results
(see, e.g., Itoh et al.\ 1992),
we take $T$ and ${\tilde \rho}$
as independent variables.
It is convenient to introduce the notation
$f \equiv \hbar \omega_{\rm p}/(k_{\rm B} T)$.

The expression for the plasma frequency in the
Braaten-Segel approximation has the form
(see Braaten $\&$ Segel 1993)
\begin{equation}
\omega_{\rm p}^2 = \frac{4\alpha}{\pi} \, {c^3 \over \hbar^2} \,
\int_0^\infty dp \,\,
\frac{p^2}{E} \left( 1-\frac{1}{3}v^2 \right)
\left[n_{\rm F}(E)+\overline{n}_{\rm F}(E) \right],
\label{wp}
\end{equation}
where $p$, $v=pc/E$, and $E=\sqrt{p^2 c^2 + m_{\rm e}^2 c^4}$
are, respectively, the momentum,
dimensionless velocity,
and energy of an electron or positron;
$n_{\rm F}(E)=1/\{{\rm exp}[(E-\mu)/(k_{\rm B} T)]+1\}$
is the Fermi-Dirac distribution for electrons;
$\overline{n}_{\rm F}(E)=1/\{{\rm exp}[(E+\mu)/(k_{\rm B} T)]+1\}$
is the Fermi-Dirac distribution for positrons;
$\mu$ is the electron chemical potential.

In the region of relativistic temperatures
($k_{\rm B} T \gg m_{\rm e} c^2$)
and under the condition $k_{\rm B} T \gg p_{\rm F} c$,
the plasma frequency (\ref{wp}) has the asymptote
\begin{equation}
\omega_{\rm p}^2 = \frac{4 \pi \alpha}{9} \;
{(k_{\rm B} T)^2 \over \hbar^2}.
\label{Wasy2}
\end{equation}
Here $p_{\rm F} \equiv (3 \pi^2 \hbar^3 n_{\rm e})^{1/3}$.
For a degenerate electron gas $p_{\rm F}$
is the usual Fermi momentum of the electrons.

In the case when ({\it i}) the electron gas is degenerate
($k_{\rm B} T \ll
\sqrt{p_{\rm F}^2 c^2 + m_{\rm e}^2 c^4} - m_{\rm e}c^2$
and the contribution of positrons
to $\omega_{\rm p}$ can be neglected),
or ({\it ii}) the gas is non-degenerate, non-relativistic,
and the temperature is not too high for the appearance of positrons
[see, e.g., Landau $\&$ Lifshitz 1980, section 105],
expression (\ref{wp}) reduces to
\begin{equation}
\omega_{\rm p}^2=\frac{4\alpha}{3\pi} \, {c^3 \over \hbar^2} \,
\frac{p_{\rm F}^3}{\sqrt{p_{\rm F}^2 c^2 + m_{\rm e}^2 c^4}}.
\label{Wasy1}
\end{equation}
If the gas is non-relativistic ($p_{\rm F} \ll m_{\rm e} c$),
then this equation gives
the well-known result,
$\omega_{\rm p}^2 = 4 \pi e^2 n_{\rm e}/m_{\rm e}$.
Notice that, since the contribution of the positrons
to the asymptote (\ref{Wasy1}) is negligible
($n_{\rm e^+} \ll n_{\rm e}$),
$p_{\rm F}$ in this case can be approximately calculated as
$p_{\rm F} \approx [3 \pi^2 \hbar^3 \, (n_{\rm e}-n_{\rm e^+})]^{1/3}
= (3 \pi^2 \hbar^3 \, \tilde{\rho}/m_{\rm u})^{1/3}$
(see equation \ref{ne}).
Introducing a new dimensionless parameter,
${\tilde p}_{\rm F}
\equiv (\hbar /m_{\rm e} c) \, (3 \pi^2 \, \tilde{\rho}/m_{\rm u})^{1/3}$,
one can substitute $(m_{\rm e} c \, {\tilde p}_{\rm F})$
for $p_{\rm F}$ in the asymptote~(\ref{Wasy1}).

Braaten $\&$ Segel (1993) developed
a useful approximate method
to calculate the emissivity due to plasmon decay.
Below in this section we present some results
obtained using this method
(more details are given
in the original paper of the authors).

Using the method of Braaten $\&$ Segel,
the emissivity can be expressed through
the parameter $v_\ast$,
which is a characteristic dimensionless velocity
of electrons scaling from 0
in the non-relativistic limit
to 1 in the ultrarelativistic limit,
\begin{equation}
v_\ast=\frac{\omega_1}{\omega_{\rm p}} \label{v_star}.
\end{equation}
Here, the plasma frequency $\omega_{\rm p}$
is given by equation (\ref{wp}) while
the frequency $\omega_1$ is
\begin{equation}
\omega_1^2=\frac{4\alpha}{\pi} \, {c^3 \over \hbar^2} \,
\int_0^\infty dp \; \frac{p^2}{E}
\left( \frac{5}{3}v^2-v^4 \right)
\left[n_{\rm F}(E)+\overline{n}_{\rm F}(E) \right].
\label{w1}
\end{equation}
In two limiting cases the neutrino emissivity due to decay
of longitudinal and transverse plasmons can be
calculated analytically.
If the plasma frequency is much smaller than the temperature
[$f \equiv \hbar \omega_{\rm p}/(k_{\rm B} T) \ll 1$],
then equations (\ref{QT}) and (\ref{QL}) can be simplified
and written~as
\begin{eqnarray}
Q_{\rm t} &=& Q_{\rm 0} \, \left({k_{\rm B}T \over m_{\rm e} c^2}\right)^9
\;\, 4 \zeta_3 \, \beta^6 \, f^6,
\label{QTasy1} \\
Q_{\rm l} &=& Q_{\rm 0} \, \left({k_{\rm B}T \over m_{\rm e} c^2}\right)^9
\; \, A(v_\ast) \, f^8.
\label{QLasy1}
\end{eqnarray}
Here, $\zeta_3\simeq 1.202057$ is a value of the Riemann zeta-function and
the function $\beta(v_\ast)$ equals
\begin{equation}
\beta=\left[ \frac{3}{2v_\ast^2}
\left( 1-\frac{1-v_\ast^2}{2v_\ast}\;
{\rm ln} \frac{1+v_\ast}{1-v_\ast} \right)  \right]^{1/2}.
\label{beta}
\end{equation}
In the non-relativistic limit ($v_{\ast} \rightarrow 0$)
it reduces to $\beta = 1$,
while in the ultra-relativistic limit
($v_{\ast} \rightarrow 1$) one has $\beta = \sqrt{3/2}$.
Furthermore, $A(v_\ast)$ is a smooth function of $v_\ast$,
changing from $8/105 \approx 0.076$ at $v_\ast \rightarrow 0$
to 0.349 at $v_\ast \rightarrow 1$.
If the plasma frequency is much greater than the temperature
($f \gg 1$), then the integrals (\ref{QT}) and (\ref{QL}) can
be taken analytically,
\begin{eqnarray}
Q_{\rm t} &=& Q_{\rm 0} \left({k_{\rm B}T \over m_{\rm e} c^2}\right)^9
\,\; b_1 \, f^{7.5} \, {\rm exp}(-f),
\label{QTasy2} \\
Q_{\rm l} &=& Q_{\rm 0} \left({k_{\rm B}T \over m_{\rm e} c^2}\right)^9
\,\; b_2 \, f^{7.5} \, {\rm exp}(-f),
\label{QLasy2}
\end{eqnarray}
where $b_1=\sqrt{2\pi}\; (1+v_\ast^2/5)^{-3/2}$ and
$b_2=\sqrt{\pi/2}\; (3v_\ast^2/5)^{-3/2}$.

\section{Fit for plasma frequency}
To simplify subsequent analysis we derived
an analytical formula which approximates the plasma frequency (\ref{wp})
in a wide range of temperatures $T=(10^7-10^{11})$ K
and effective densities
${\tilde \rho} = (2 \times 10^2-10^{14})$ g cm$^{-3}$.
This range of parameters includes
all possible limiting cases of
degenerate, ultrarelativistic,
as well as of non-degenerate non-relativistic electron gas.
We calculated the emissivity on a dense grid of mesh points
(with the steps 0.2 in ${\rm lg} \, T$ and ${\rm lg} \, \tilde{\rho}$).
The root mean-square relative error of our approximation is 0.4\%.
The maximum error of 1.4\% is at
${\rm lg} \, T=9.0$ (K) and
${\rm lg} \, {\tilde \rho}=2.4$ (g cm$^{-3}$).
The fit reproduces the asymptotes from Section 2.
The squared plasma frequency can be approximated as
\begin{equation}
\omega_{\rm p}^{2} =
\left( \frac{m_{\rm e} \, c^2}{\hbar} \right)^2
\sqrt{ {\rm asy}_2^{2}+[{\rm asy}_1 \; (1-C \; D)]^2 }.
\label{wpfit}
\end{equation}
Here, ${\rm asy}_1 = 4 \alpha /(3\pi) \;
\tilde{p}_{\rm F}^3 / \sqrt{1+\tilde{p}_{\rm F}^2}$
is exactly the low-temperature asymptote (\ref{Wasy1})
[we recall that $\tilde{p}_{\rm F} = (\hbar/m_{\rm e} c) \;
(3 \pi^2 \tilde{\rho}/m_{\rm u})^{1/3}$],
while ${\rm asy}_2$ is given by
\begin{equation}
{\rm asy}_2=\frac{4 \pi \alpha}{9} p_2
\left( \frac{t^2}{p_2}+1+\frac{p_2}{t^2} \right)
\left[1+\frac{p_3}{(t/\sqrt{p_2})^{p_1}}\right]^{-10},
\label{asy2}
\end{equation}
with $t \equiv k_{\rm B} T /(m_{\rm e} c^2)$.
In the high-temperature limit, $\rm{asy_2}$
transforms into the asymptote (\ref{Wasy2}).
The fit parameters $p_1$, $p_2$, and $p_3$ equal
$p_1= 1.793$, $p_2= 0.0645$, and $p_3= 0.433$.

The function $C$ in equation (\ref{wpfit}) is written as
\begin{equation}
C=1-c_2\; \frac{(c_1\; t)^2}{1+(c_1\; t)^2} \,,
\label{AA}
\end{equation}
where
\begin{eqnarray}
c_1 &=& p_4 \; \frac{1+p_5\; \tilde{\rho}^{p_6}}{1+p_7 \;
(1+p_5\; \tilde{\rho}^{p_6})} \,,
\label{a1} \\
c_2 &=& p_8+p_9\frac{\tilde{\rho}}{p_{10}+\tilde{\rho}}\,,
\label{a2}
\end{eqnarray}
%
with
$p_4 = 0.01139$, $p_5 = 2.484 \times 10^6$,
$p_6= -0.6195$, $p_7= 0.0009632$,
$p_8= 0.4372$, $p_9= 1.614$,
and $p_{10}=8.504\times 10^8$.

At low temperatures the plasma frequency
in the first approximation depends only on
${\tilde \rho}$
and we have $C=1$.
The function $D$ in equation (\ref{wpfit}) has the form
\begin{eqnarray}
D     &=& \frac{t^2}{d_{1} \sqrt{1+(d_{2}\; t)^2}},
\label{D} \\
d_{1} &=& \frac{6}{\pi^2} \,\,
\frac{\tilde{p}_{\rm F}^2 (1 + \tilde{p}_{\rm F}^2) }
{2 \tilde{p}_{\rm F}^2 + 5}, \qquad
d_{2} = \frac{\pi^2}{6} \,\,
\frac{1}{\sqrt{1+\tilde{p}_{\rm F}^2}-1}.
\label{d1d2}
\end{eqnarray}
At high temperatures
(when $t \gg 1$ and the electron gas is non-degenerate)
the fit (\ref{wpfit}) reproduces
the high-temperature asymptote (\ref{Wasy2}).
At low temperatures
(a degenerate gas or a non-degenerate non-relativistic gas;
positrons can be neglected) the fit (\ref{wpfit})
transforms into the analytical asymptote (\ref{Wasy1}),
which depends only on $\tilde{\rho}$.
The function $D$ is designed in such a way
to reproduce not only the asymptote (\ref{Wasy1})
of plasma frequency
but also the first temperature corrections
to $\omega_{\rm p}$.
For a degenerate electron gas,
the expansion parameter is $k_{\rm B}T/\mu$;
for the non-degenerate non-relativistic gas it reduces to
$k_{\rm B}T/(m_{\rm e}c^2)$.

\section{Fit for the neutrino emissivity}
In this section we present an analytical formula
which approximates the results of numerical calculations
of the emissivity $Q=Q_{\rm t}+Q_{\rm l}$ (per unit volume) and reproduces
the asymptotes from Section 2.
The approximation was made in a range of temperatures
$T=(10^7-10^{11}$)~K and effective densities
$\tilde{\rho}=(2\times10^2-10^{14}$)~g~cm$^{-3}$.
The emissivity $Q({\tilde \rho}, T)$ was calculated on the same
grid points as the plasma frequency (Section 3).
At $f \equiv \hbar \omega_{\rm p}/(k_{\rm B} T) >20$
the accuracy of our fit is only logarithmic.
However, in this case the emissivity $Q$ is
exponentially small, $Q \sim {\rm exp}(-f)$.

The fit for the emissivity can be presented in the form
\begin{equation}
Q = Q_{\rm l} + Q_{\rm t} =
Q_{\rm 0} t^9 \, (W_{\rm t}+W_{\rm l}) \; {\rm exp}(-f),
\label{Qfit}
\end{equation}
where, as before, $t\equiv k_{\rm B}T/(m_{\rm e}c^2)$
and we define
\begin{eqnarray}
W_{\rm t} &\equiv& {\rm asy}_{\rm t1} + {\rm asy}_{\rm t2}\;
{\rm exp}\left[\frac{q_3}{(f^{q_1}+ q_2)}\right],
\label{Qt22} \\
W_{\rm l} &\equiv& \frac{{\rm asy}_{\rm l2}\; [{\rm asy}_{\rm l1}+q_4 \;
(1+q_5 \; v_\ast^{2.5})^{3.5} f^9]}
{{\rm asy}_{\rm l2}+[{\rm asy}_{\rm l1}+q_4 \;
(1+q_5 \; v_\ast^{2.5})^{3.5} f^9]}.
\label{Ql22}
\end{eqnarray}
In equations (\ref{Qt22}) and (\ref{Ql22})
\begin{eqnarray}
{\rm asy}_{\rm t1} &=& a_1 f^6, \qquad \;\,
{\rm asy}_{\rm l1} = a_2 f^8 ,
\label{asyt1l1} \\
{\rm asy}_{\rm t2} &=& b_1 f^{7.5}, \qquad
{\rm asy}_{\rm l2} = b_2 f^{7.5},
\label{asyt2l2} \\
a_1 &=& 4 \zeta_3 \beta^6,  \qquad \;\,\,
\,\,\, a_2 = \frac{8}{105} +
\left(0.349-\frac{8}{105}\right) v_\ast^{10};
\end{eqnarray}
the functions $b_1(v_\ast)$ and $b_2(v_\ast)$
are the same as in equations (\ref{QTasy2}) and (\ref{QLasy2});
the function $\beta(v_\ast)$ is given by equation (\ref{beta}).
At $f\ll 1$ equation (\ref{Qfit}) transforms into
\begin{equation}
Q = Q_{\rm t} = Q_{\rm 0} t^9 {\rm asy}_{\rm t1}
= Q_{\rm 0} t^9 \; 4 \zeta_3 \beta^6 f^6
\label{Qasy1}
\end{equation}
(compare with the asymptotes \ref{QTasy1} and \ref{QLasy1}).
At $f\gg 1$ one has
\begin{equation}
Q = Q_0 t^9 ({\rm asy}_{t2}+{\rm asy}_{l2})\; {\rm exp}(-f)
= Q_0 t^9 (b_1+b_2)f^{7.5} \; {\rm exp}(-f)
\label{Qasy2}
\end{equation}
(compare with the asymptotes \ref{QTasy2} and \ref{QLasy2}).

When calculating the emissivity
from equation (\ref{Qfit})
one should use the fit (\ref{wpfit})
for the plasma frequency $\omega_{\rm p}$
and the following fit
for the characteristic velocity $v_{\ast}$,
\begin{eqnarray}
v_\ast=\left(\frac{{\tilde v}_{\rm F}^3 + s_1 \;
t^{s_2} \tilde{\rho}^{s_3}}
{1 + s_1 \; t^{s_2} \tilde{\rho}^{s_3}}\right)^{1/3},
\label{vfit}
\end{eqnarray}
where ${\tilde v}_{\rm F} \equiv
{\tilde p}_{\rm F}/\sqrt{1+{\tilde p}_{\rm F}^2}$;
$s_1=  9.079$; $s_2=  1.399$; $s_3= -0.06592$.
The root mean-square relative error of this approximate formula
in the chosen range of $T$ and ${\tilde \rho}$
constitutes 1.4\%.
The maximum fit error is equal to 5.4\%
at ${\rm lg} \, T=8.4$ (K) and
${\rm lg} \, {\tilde \rho}=3.8$ (g cm$^{-3}$).

In addition, it turns out to be necessary
to use a special approximate formula for the function
$\beta^6(v_\ast)$ from which
the fitting expression (\ref{Qfit}) depends on
[a simple substitution of equation (\ref{vfit}) into (\ref{beta}) and
subsequent calculation of $\beta^6$ results in large errors],
\begin{eqnarray}
\beta^6=\beta^6({\tilde v}_{\rm F})
+ \left[ 3.375 - \beta^6({\tilde v}_{\rm F}) \right] \frac{t^{r_2}
\tilde{\rho}^{r_3}}{( r_1 + t^{r_2} \tilde{\rho}^{r_3} )},
\label{beta6}
\end{eqnarray}
where $r_1=  0.3520$; $r_2= 1.195$; $r_3= -0.1060$.
The root mean-square relative error of this fit constitutes 2.5\%,
the maximum error of 8.3\%
is at ${\rm lg} \, T=9.6$ (K) and
${\rm lg} \, {\tilde \rho}=5.0$ (g~cm$^{-3}$).
The function (\ref{beta6}) was approximated
in the same temperature and density range
as the parameter $v_\ast$ and the emissivity $Q$.

%
%
\begin{figure}
\begin{center}
\leavevmode
\epsfxsize=12cm
\epsfbox[15 80 540 560]{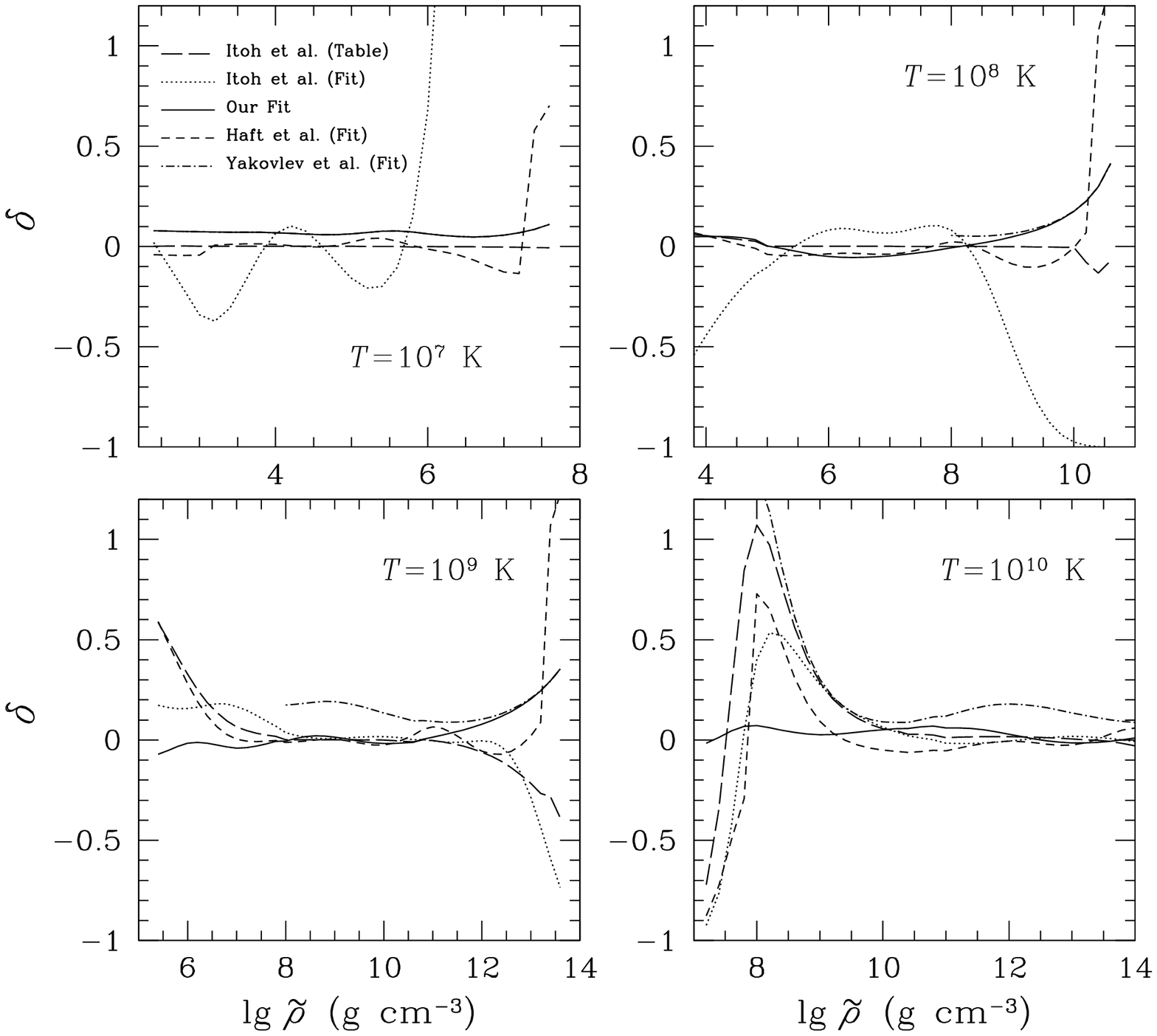}
\end{center}
\caption{
Relative deviation
$\delta \equiv (Q_{\rm lit}-Q_{\rm num})/Q_{\rm num}$
versus $\tilde{\rho}$
for $T=10^7$, $10^8$, $10^9$, and $10^{10}$~K.
Here, $Q_{\rm num}$ is the emissivity,
numerically calculated in this paper.
For $Q_{\rm lit}$ we take one of the emissivities
obtained either from the fit formula (\ref{Qfit}) (solid lines);
or from the table of Itoh et al.\ (1992) (long dashes);
or from the fit formula of Itoh et al.\ (1992) (dots);
or from the fit formula of Haft et al.\ (1994) (short dashes);
or from the approximate formula from the review of
Yakovlev et al.\ (2001) (dot-dashed lines).}
\label{fig:fig2}
\end{figure}
%
%
The use of approximate formulae
(\ref{wpfit}) and (\ref{vfit})--(\ref{beta6})
leads to the following values of fitting parameters
$q_1, \ldots, q_5$ (see equations \ref{Qt22} and \ref{Ql22}),
minimizing root mean-square deviation
of the emissivity, provided by equation (\ref{Qfit}),
from the numerical values,
\begin{eqnarray}
q_1 &=&  0.7886,\; q_2= 0.2642,
\nonumber \\
q_3 &=& 1.024, \; q_4=  0.07839,\; q_5 = 0.1784.
\label{approx_q1}
\end{eqnarray}
The root mean-square relative error
of the approximate formula (\ref{Qfit})
with these coefficients is 4\%,
the maximum error is 7.9\% at
${\rm lg} \, T=8.4$ and
${\rm lg}\, \tilde{\rho}=6.4$.

In Fig.\ 2 we compare
our numerical results for the emissivity $Q_{\rm num}$
with the results taken from the literature
(corresponding emissivities are denoted as $Q_{\rm lit}$).
The figure presents the relative deviation
$\delta \equiv (Q_{\rm lit}-Q_{\rm num})/Q_{\rm num}$
as a function of $\tilde{\rho}$
for a set of temperatures
$T=10^7$, $10^8$, $10^9$, and $10^{10}$~K.
The solid curves demonstrate relative deviations
of the approximation (\ref{Qfit}), suggested in this paper,
from our numerical results $Q_{\rm num}$;
the long dashes show deviations
from numerical calculations of Itoh et al.\ (1992)
[taken from their table];
the dotted curves correspond to
relative deviations calculated
using an approximate formula,
suggested by Itoh et al.\ (1992);
the short dashes describe relative deviations
calculated from a fitting formula of Haft et al.\ (1994).
Finally, by the dot-dashed curves
we show relative deviations
calculated from the approximate formula for the emissivity
given in the review of Yakovlev et al.\ (2001).
In that review it is recommended
to use the formula
only for $\tilde{\rho}>10^8$ g cm$^{-3}$
and for strongly degenerate electrons.
From the analysis of Fig.\ 2
a number of conclusions
can be inferred:

({\it i}) The approximate formula obtained in this section
is in good agreement with the results
of numerical calculations as long as $f<20$
(at greater $f$, that is at higher densities,
the solid curve tends to go upward).

({\it ii}) Our calculations agree with results of Itoh et al.\ (1992)
in the range of parameters,
where the electron gas is strongly degenerate
and the emissivity is not small.
However, as follows, for example,
from Fig.\ 2 at $T=10^9$~K,
some our results deviate from those of Itoh et al.\ (1992) for
$\tilde{\rho} \sim 10^{13}$ g cm$^{-3}$.
For this case, matter is strongly degenerate
so that the simplified assumptions, made by Itoh et al.\
at calculating the emissivity, could not lead
to such deviations.
(Let us note that Itoh et al. used
the dielectric function, calculated by Jancovici 1962
for a strongly degenerate electron gas, see Appendix A.)
Taking into account that our numerical results
at such densities and $T=10^9$~K
do not differ from the analytical asymptote for the emissivity
by more than 10\%,
the results of Itoh et al.\ (1992)
in the indicated parameter range seem less accurate than ours.

({\it iii}) The fit formula of Itoh et al.\ (1992)
satisfactorily describes the results
of numerical calculations
only near the maximum of the emissivity
(when $f \sim 1$).

({\it iv}) The fit formula of Haft et al.\ (1994)
agrees well with our numerical results
in the same region of temperatures and densities
in which the numerical results of Itoh et al.\ (1992) agrees with
our numerical results.

({\it v}) The approximate formula
from the review of Yakovlev et al.\ (2001)
becomes inaccurate
at high temperatures (i.e., $T=10^{10}$~K)
and low densities (${\tilde \rho} \sim 10^8$ g cm$^{-3}$).
This approximate formula is valid only for
strongly degenerate electrons,
while the electron degeneracy becomes mild
at high $T$ and low ${\tilde \rho}$.

Summarizing, as follows from Fig.\ 2, the results
of various authors are in {\it satisfactory} agreement
in the ranges of $T$ and ${\tilde \rho}$
where the process of neutrino emission
due to plasmon decay is the most efficient mechanism
of energy losses in dense stellar matter.

\section{The neutrino luminosity of white dwarfs}

Let us apply the results of Section 4
to analyze the neutrino luminosity of white dwarfs.
As will be argued below,
the neutrino luminosity
due to plasmon decay only weakly depends on
a specific model of a white dwarf.
Thus, it can be considered
as a universal function
of the white dwarf mass $M$
and its internal temperature $T$.
Here we calculate this
universal function
and approximate it
by a convenient analytical formula.

As is well known, the thermal evolution of a white dwarf consists of
two stages, the neutrino cooling stage (where cooling is mainly
realized through the neutrino emission from the entire stellar body)
and the photon stage (the main energy losses through the photon
radiation from the stellar surface). A transition from one stage to
the other occurs at the stellar age $\tau \sim (10^7-10^8)$ yr, when
the surface temperature of a star equals $T_{\rm s} \sim 2.5 \times
10^4$~K (for a hydrogen or helium atmosphere white dwarf,
see, e.g., Winget et al.\ 2004).

At the neutrino cooling stage the main mechanism of energy losses
is the neutrino emission due to plasmon decay
(the second important process -- the neutrino bremsstrahlung
in collisions of electrons with atomic nuclei --
is 10--100 times weaker, see Winget et al.\ 2004).
We numerically calculated the neutrino luminosity $L_\nu(M, T)$
of white dwarfs caused by the decay of plasmons.
When doing the calculation,
we made the following assumptions.
First, to obtain the density profile inside a white dwarf
we assumed that the pressure is fully determined
by degenerate electrons.
Second, the stellar core was assumed to be isothermal,
which is a good approximation
for not too young white dwarfs
($\tau \ga 10-1000$ yr).
Third, we neglected beta-captures
when calculating the structure and luminosity
of massive white dwarfs.
Beta-captures lead to softening of the equation of state,
and influence the hydrostatic structure of a star.
In addition, they change stellar
chemical composition,
affect the number of nucleons per one electron,
$\mu_{\rm e}$, and, consequently, the quantities
${\tilde \rho}$ and $L_{\nu}$.
However, because the neutrino luminosity
is the integral characteristic of a star,
it should not strongly depend
on these simplified assumptions.

In Fig.\ 3 we present the neutrino luminosity $L_\nu$
as a function of stellar core temperature $T$ for white dwarfs
with the masses $M=0.4M_\odot$, $0.6M_\odot$,
$M_\odot$, and $1.4M_\odot$.

\begin{figure}
\begin{center}
\leavevmode
\epsfxsize=8cm
\epsfbox[60 200 560 670]{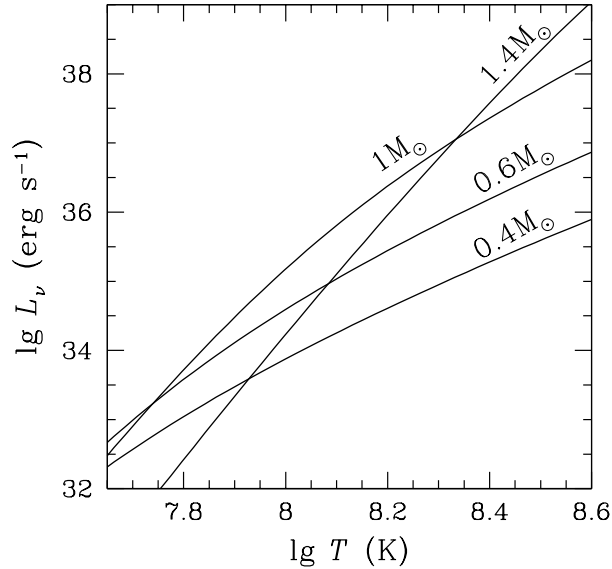}
\end{center}
\caption{
The neutrino luminosity $L_{\nu}$ versus
internal stellar temperature $T$ for white dwarfs with
$M=0.4M_\odot$, $0.6M_\odot$,
$M_\odot$, and $1.4M_\odot$.}
\label{fig:fig3}
\end{figure}
%
The results of numerical calculations
of $L_{\nu}$ in the range of temperatures
$T=(3\times10^7-5\times10^8)$~K
and masses $M=(0.4-1.3)M_\odot$
were approximated by the formula
\begin{equation}
L_{\nu 1}(M,T) = 10^{39} \,\, \frac{k_1 T_8^{31/3} (k_4 \widetilde{M}^{k_2}
+ \widetilde{M}^{k_3})(1+k_5 \widetilde{M})^{22/3}}
{\left[ k_6(1+k_5 \widetilde{M})^{22/(3k_7)}T_8^{22/(3k_7)}
+ \widetilde{M}^{22/(3k_7)}\right]^{k_7}} \quad \rm{erg~ s^{-1}},
\label{Lfit}
\end{equation}
where $\widetilde{M}=M/M_{\odot}$,
$T_8=T/(10^8 {\rm K})$, and
%
\begin{eqnarray}
&&k_1 = 1.050, \;
k_2=11.86, \; k_3=5.901,
\nonumber\\
&&k_4 = 1.010, \; k_5=-0.5448, \; k_6=2.777, \; k_7=5.635.
\label{ki}
\end{eqnarray}
For white dwarfs with $M=(1.3 - 1.4) M_\odot$
the neutrino luminosity in the same range of
temperatures $T=(3\times10^7-5\times10^8)$~K
is given by
\begin{equation}
L_{\nu 2}(M,T) = 10^{39} \,\, \frac{l_1 T_8^{31/3} \widetilde{M}^{l_2}}
{\left(l_3 T_8^{22/(3l_5)} +
\widetilde{M}^{22l_4/(3l_5)}\right)^{l_5}}
\quad \rm{erg~s^{-1}},
\label{LfitMas}
\end{equation}
where
%
\begin{equation}
l_1=2.777, \;
l_2=25.13, \; l_3=3.095, \;
l_4=7.585, \; l_5=7.381.
\label{li}
\end{equation}
The maximum error of the fit expressions
(\ref{Lfit}) and (\ref{LfitMas})
does not exceed 14\%.
Unfortunately, these two approximations
do not match at $M=1.3M_\odot$.
Thus, to calculate the neutrino emissivity
for a white dwarf with the mass
$M \in [1.28M_\odot, 1.32M_\odot]$,
we recommend to use a linear interpolation
\begin{eqnarray}
L_{\nu 3}(M,T) = L_{\nu 1}(1.28M_\odot,T)
+ \frac{L_{\nu 2}(1.32M_\odot,T)
-L_{\nu 1}(1.28 M_\odot,T)}{0.04}(\widetilde{M}-1.28).
\label{sew}
\end{eqnarray}
This interpolation does not affect
the maximum fit error which remains to be 14\%
at $T=2.38 \times 10^8$ K and $M=1.34 M_{\odot}$.

As seen from equations
(\ref{Lfit}) and (\ref{LfitMas}),
in the limit of high temperatures
$L_{\nu} \sim T^3$, while in the limit
of low temperatures $L_{\nu} \sim T^{31/3}$.
Let us demonstrate how to obtain
this temperature dependence
from simple physical arguments.

At high temperatures,
the internal stellar temperature $T$ is much greater than
the plasma frequency $\omega_{\rm p0}$
in the center of the star.
Since the plasma frequency of degenerate electrons
becomes smaller as the density decreases
(see equation \ref{Wasy1}), we have
$k_{\rm B} T \gg \hbar \omega_{\rm p}$ throughout the star.
In this case the neutrino emissivity
of an arbitrary volume element in the star
is given by asymptote (\ref{Qasy1})
and the luminosity equals
\begin{equation}
L_\nu \approx 4 \xi_3 \,
\frac{\hbar^6 k_{\rm B}^3}{(m_{\rm e}c^2)^9} \, \,
Q_{\rm 0} T^3 \,
\int_{\rm star} \beta^6 \,
\omega_{\rm p}^6 \,\, {\rm d} V.
\label{lumi1}
\end{equation}
Here the integral is taken over the volume $V$ of the star.
Since the plasma frequency $\omega_{\rm p}$  and
the parameter $\beta$ depend only on $\tilde{\rho}$
(see equations \ref{Wasy1} and \ref{beta}),
one gets~$L_\nu\propto T^3$.

In the low-temperature limit,
when $k_{\rm B} T \ll \hbar\omega_{\rm p0}$,
the main contribution to the luminosity comes from
a thin spherical layer of width $h$,
in which $\hbar \omega_{\rm p} \sim~k_{\rm B}T$.
This layer is situated in the outermost part of the stellar core,
where the electrons form a degenerate, non-relativistic gas.
Indeed, if we move from this layer to the stellar center,
$\omega_{\rm p}$ will increase
while the emissivity will be exponentially suppressed,
$Q \sim {\rm exp}(-\hbar \omega_{\rm p}/k_{\rm B}T)$,
in accordance with equation (\ref{Qasy2}).
If we move from the layer to the stellar surface
then the emissivity will also decrease
(see asymptote \ref{Qasy1}) but in a power-law fashion,
$Q \sim \beta^6 \omega_{\rm p}^6 = \omega_{\rm p}^6$
($\beta=1$ for the non-relativistic electron gas,
see equation \ref{beta}).
Therefore, the emissivity will have a maximum
in a layer in which
$\hbar \omega_{\rm p} \sim~k_{\rm B}T$,
and the neutrino luminosity of a star
can be estimated as
\begin{equation}
L_{\nu} \sim 4 \xi_3 \beta^6 \, Q_{\rm 0} \,
\left( \frac{k_{\rm B} T}{m_{\rm e}c^2} \right)^9
\left(\frac{\hbar \omega_{\rm p}}{k_{\rm B}T}
\right)^6 4 \pi R^2 h
\sim 16 \pi \xi_3 \, Q_{\rm 0}
\left(\frac{k_{\rm B}T}{m_{\rm e} c^2} \right)^9 R^2 h,
\label{lumi2}
\end{equation}
where $R$ is the white dwarf radius.
An order of magnitude estimate gives
the characteristic width $h$ of the layer,
$h \sim \omega_{\rm p}^6
/({\rm d}\omega_{\rm p}^6/{\rm d}r)$.
Using the hydrostatic equilibrium equation
and the scaling relations for the plasma frequency
(see equation \ref{Wasy1})
$\omega_{\rm p} \propto \tilde{\rho}^{1/2}$
and pressure $P \propto \tilde{\rho}^{5/3}$
of the degenerate non-relativistic gas,
we get $h \propto \tilde{\rho}^{2/3}
\propto \omega_{\rm p}^{4/3} \propto T^{4/3}$.
Consequently, $L_{\nu} \propto T^{31/3}$,
in agreement with the estimate (\ref{lumi2}).


Let us note that the plasmon decay
neutrino emissivity and hence the luminosity
of the star depend on the effective density $\tilde{\rho}$,
which is related to the real density $\rho$
by equation (\ref{rho}),
$\tilde{\rho}/\rho
=1/\mu_{\rm e}
=\sum_i Z_i n_i/\left(\sum_i A_i n_i \right)$.
In white dwarfs with any reasonable chemical composition,
the mass number $A_i$ of atomic nuclei species $i$ is
always twice as much than the charge number $Z_i$
(recall that we neglect beta-captures).
Thus, the ratio $\tilde{\rho}/\rho$ is equal to 1/2.
We used this ratio in all our calculations.

\section{Summary}

We have calculated the neutrino emissivity
$Q$ due to plasmon decay in an electron-positron plasma
making no assumptions about degree of degeneracy
or relativity of the electron gas.

When calculating the emissivity
one needs
the plasma dielectric functions as well as
the dispersion relations for transverse and longitudinal
plasmons in a wide range of temperatures and densities.
In particular, we have calculated the plasma frequency
$\omega_{\rm p}$ and fitted it by
an analytical formula.
This formula reproduces
the main asymptotes for $\omega_{\rm p}$
(degenerate, ultrarelativistic
or non-degenerate non-relativistic
electrons, see Section~3).

The results of numerical calculations of the neutrino emissivity
were also approximated by a convenient analytical expression.
It satisfies the asymptotes in various limiting cases
(Section 4, also see the paper by Braaten $\&$ Segel 1993).
The approximation is valid for
$T=(10^7-10^{11})$~K and
$\tilde{\rho}=(2\times 10^2-10^{14})$~g~cm$^{-3}$.
The root mean-square relative error
of the approximation does not exceed 4\%
for those temperatures and densities, for which
$f=\hbar \omega_{\rm p}/(k_{\rm B}T)<20$
[while at $f>20$ the emissivity is exponentially small,
$Q \sim {\rm exp}(-f)$].

The fit expression for the emissivity was used
to calculate the neutrino luminosity of white dwarfs
(Section 5).
This neutrino luminosity was fitted
by analytic formulas and presented as a function of
white dwarf mass and its internal temperature.
It is shown that the neutrino luminosity depends
on the chemical composition of a white dwarf
only through the parameter $\mu_{\rm e}$
which is equal to 2 for reasonable white dwarf compositions.

The results of this paper can be used in a number
of applications, in particular, in modelling of
the evolution of red giants or presupernovae as well as
in the cooling theory of white dwarfs
(see, e.g., Haft et al.\ 1994, Winget et al.\ 2004).

\section*{Acknowledgments}

The authors are grateful
to D.G. Yakovlev for discussions;
to our referee, Agnes Kim, for valuable suggestions and comments;
to A.I. Chugunov for providing the code
that was used to approximate numerical
results by analytical functions;
to A.Y. Potekhin
for calculating the relation
between the surface and internal temperatures
of helium atmosphere white dwarfs;
and to D.P. Barsukov
and A.M. Krassilchtchikov
for technical assistance.
This research was supported by RFBR
(grants 05-02-16245 and 05-02-22003)
and by the Federal Agency
for Science and Innovations
(grant NSh 9879.2006.2).


\appendix

\section[]{Dielectric functions of electron-positron plasma}
\label{appendixA}

Using the density matrix formalism
we calculated the dielectric function
of the electron-positron gas in the first order
of perturbation theory.
The longitudinal $\varepsilon_{\rm l}(\omega, k)$
and transverse $\varepsilon_{\rm t}(\omega, k)$
components of the dielectric tensor can be written in the form
($c=\hbar=k_{\rm B}=1$)
\begin{eqnarray}
\varepsilon_{\rm l} &=&
1 - \frac{4 \pi {\rm \alpha}}{\omega^2} \,
\sum_{e^-, e^+} \,
\int \frac{{\rm d}^3 {\pmb p}}{(2 \pi)^3}
\, \, \frac{1}{E_{{\pmb p}+{\pmb k}} E_{\pmb p}} \, \,
\frac{n_{{\pmb p}+{\pmb k}} -n_{\pmb p}}
{E_{{\pmb p}+{\pmb k}} - E_{\pmb p} - \omega - i \delta} \,
\nonumber \\
&\times& \left[ 2 \frac{({{\pmb p} \cdot {\pmb k}})^2}{k^2}
+ ({{\pmb p} \cdot {\pmb k}})
+ E_{{\pmb p}+{\pmb k}} E_{\pmb p} - E_{\pmb p}^2 \right],
\label{epsl}\\
\varepsilon_{\rm t} &=& 1 - \frac{4 \pi {\rm \alpha}}{\omega^2} \,
\sum_{e^-, e^+} \,
\int \frac{{\rm d}^3 {\pmb p}}{(2 \pi)^3}
\, \, \frac{1}{E_{{\pmb p}+{\pmb k}} E_{\pmb p}} \, \,
\frac{n_{{\pmb p}+{\pmb k}} -n_{\pmb p}}
{E_{{\pmb p}+{\pmb k}} - E_{\pmb p} - \omega - i \delta} \,
\nonumber \\
&\times& \left[ \frac{( {\pmb p} {\pmb \times} {\pmb k})^2}{k^2}
- ({{\pmb p} \cdot {\pmb k}})
+ E_{{\pmb p}+{\pmb k}} E_{\pmb p} - E_{\pmb p}^2 \right].
\label{epstr}
\end{eqnarray}
Here, the summation is carried over electrons and positrons;
$n_{{\pmb {\rm p}}}=1/[{\rm exp}
(E_{\pmb p} \mp \mu)/T +1]$
is the Fermi-Dirac distribution function for electrons
(in this case one have to choose the sign --)
or positrons (the sign +);
$E_{\pmb p}=\sqrt{{\pmb p}^2 +m_{\rm e}^2}$ and
$E_{{\pmb p} + {\pmb k}}
=\sqrt{({\pmb p} + {\pmb k})^2  +m_{\rm e}^2}$ is the energy
of an electron or a positron with the momentum
${\pmb p}$ and ${\pmb p} + {\pmb k}$, respectively.

We have checked that equations (\ref{epsl}) and (\ref{epstr})
are equivalent to corresponding expressions for
the dielectric function which can be obtained from the polarization
tensor $\Pi^{\mu \nu}$ of Braaten and Segel (1993)
[see their equation A1].

The integration over the angles in equations
(\ref{epsl}) and (\ref{epstr}) can be done analytically.
As a result, one obtains for real parts of
$\varepsilon_{\rm l}$ and $\varepsilon_{\rm t}$,
\begin{eqnarray}
\varepsilon_{\rm l} &=& 1 - \frac{{\rm \alpha}}
{\pi \omega^2} \, \int_{0}^{\infty}
{\rm d} p \,\, p^2 \,
\left[ n_{\rm F}(E_p)
+ \overline n_{\rm F}(E_p) \right] \, R_{\rm l},
\label{epsrl} \\
\varepsilon_{\rm t} &=&
1 - \frac{{\rm \alpha}}{\pi \omega^2} \, \int_{0}^{\infty}
{\rm d} p \,\, p^2 \, \left[ n_{\rm F}(E_p)
+ \overline n_{\rm F}(E_p) \right] \, R_{\rm t},
\label{epsrtr}
\end{eqnarray}
where
\begin{eqnarray}
R_{\rm l} &=& - \frac{4 \omega^2}{E_p k^2}
\nonumber\\
&+& \frac{\omega^2}{2 E_p k^3 p}
\left[ (2 E_p +\omega)^2 - k^2 \right]
\ln \left| \frac{E_{p-k}^2 -
(E_p + \omega)^2}{E_{p+k}^2-(E_p+\omega)^2}  \right|
\nonumber \\
&+&  \frac{\omega^2}{2 E_p k^3 p}
\left[ (2 E_p - \omega)^2 - k^2 \right]
\ln \left| \frac{E_{p-k}^2 - (E_p - \omega)^2}{E_{p+k}^2
- (E_p - \omega)^2}  \right|,
\label{rl}\\
R_{\rm t} &=& \frac{ 2 \, (\omega^2+k^2)}{E_p k^2}
\nonumber \\
&+& \frac{1}{4 E_p k^3 p}
\left[ k^4 + 4 k^2 p^2 + 4 k^2 E_p \omega
- \omega^2 (2 E_p + \omega)^2 \right]
\ln \left| \frac{E_{p-k}^2 - (E_p + \omega)^2}{E_{p+k}^2-(E_p+\omega)^2}
\right|
\nonumber \\
&+&  \frac{1}{4 E_p k^3 p}
\left[k^4 + 4 k^2 p^2 - 4 k^2 E_p \omega
- \omega^2 (2 E_p - \omega)^2 \right]
\ln \left| \frac{E_{p-k}^2 - (E_p - \omega)^2}{E_{p+k}^2
- (E_p - \omega)^2}  \right|.
\label{rtr}
\end{eqnarray}
In equations (\ref{epsrl})--(\ref{rtr})
$n_{\rm F}(E_p)$ and $\overline n_{\rm F}(E_p)$
are the Fermi-Dirac distribution functions
for electrons and positrons, respectively;
$E_{p \pm k}= \sqrt{(p \pm k)^2+m_{\rm e}^2}$
is the energy of an electron or a positron with the
absolute value of momentum equal to $(p \pm k)$.

Knowing the dielectric functions,
the plasmon dispersion relations
can be found from the equations
\begin{equation}
\varepsilon_{\rm l}(\omega, k)=0, \qquad
\omega^2 \, \varepsilon_{\rm t}(\omega, k) = k^2.
\label{disprel}
\end{equation}

If the electron gas is completely degenerate ($T=0$),
then the integrals in equations (\ref{epsrl}) and (\ref{epsrtr})
can be taken analytically.
The result is
\begin{eqnarray}
\varepsilon_{\rm l} &=& 1 - \frac{\rm \alpha}{\pi}
\, \, \left\{ \, \,
- \frac{8}{3} \, \frac{1}{k^2} \, p_{\rm F} \sqrt{p_{\rm F}^2 + m_{\rm e}^2}
\, + \frac{2}{3} \, \sinh^{-1} \frac{p_{\rm F}}{m_{\rm e}} \right.
\nonumber \\
&+& \left. \frac{1}{3} \,
\frac{(k^2 - \omega^2 - 2 m_{\rm e}^2)}{(k^2-\omega^2)}
\sqrt{  \frac{k^2 - \omega^2 + 4 m_{\rm e}^2}{\omega^2 - k^2} }
\right.
\, L_1
\nonumber \\
&-& \left. \frac{1}{6 k^3} \, \sqrt{p_{\rm F}^2+m_{\rm e}^2} \,
\left(3 \omega^2 -3 k^2 + 4 p_{\rm F}^2 + 4 m_{\rm e}^2 \right)
\, L_2  \right.
\nonumber \\
&+& \left. \frac{\omega}{12 k^3} \,
\left(\omega^2 - 3 k^2 + 12 p_{\rm F}^2 + 12 m_{\rm e}^2 \right)
\, L_3
\, \, \right\},
\label{epsldeg}
\end{eqnarray}
\begin{eqnarray}
\varepsilon_{\rm t} &=& 1 - \frac{\rm \alpha}{\pi \omega^2}
\, \, \left\{ \, \, \frac{2}{3} \,
\frac{(k^2+2 \omega^2)}{k^2} \, p_{\rm F} \sqrt{p_{\rm F}^2 + m_{\rm e}^2}
\right.
\nonumber \\
&-& \left. \frac{2}{3} \, (k^2 - \omega^2) \, \sinh^{-1} \frac{p_{\rm F}}{m_{\rm
e}} \right.
\nonumber \\
&-& \left. \frac{(k^2 - \omega^2 - 2 m_{\rm e}^2)}{3} \,
\sqrt{  \frac{k^2 - \omega^2 + 4 m_{\rm e}^2}{\omega^2 - k^2} }
\, L_1 \right.
\nonumber \\
&+& \left. \frac{\sqrt{p_{\rm F}^2 + m_{\rm e}^2}}{k^3}
\left[ -\frac{1}{3}(k^2-\omega^2)(p_{\rm F}^2+m_{\rm e}^2)+
\frac{1}{4} \left(-k^4+\omega^4+4 m_{\rm e}^2 k^2 \right)
\right] \, L_2 \right.
\nonumber \\
&+& \left. \frac{\omega}{24 k^3}
\left[ (k^2 - \omega^2) \left(3 k^2 + \omega^2 + 12 p_{\rm F}^2 + 12 m_{\rm e}^2
\right)
- 12 m_{\rm e}^2 k^2
\right] \, L_3 \, \, \right\}.
\label{epsdegtr}
\end{eqnarray}
The quantities $L_2$ and $L_3$ are
\begin{eqnarray}
L_2 &=& \ln \left|
\frac{(-k^2 + \omega^2 - 2 k p_{\rm F})^2 - 4 \omega^2 (p_{\rm F}^2+m_{\rm
e}^2)}
{(-k^2 + \omega^2 + 2 k p_{\rm F})^2 - 4 \omega^2 (p_{\rm F}^2+m_{\rm e}^2)}
\right|,
\label{l2} \\
L_3 &=& \ln \left|
\frac{(-k^2+\omega^2)^2
- 4 (\omega \sqrt{p_{\rm F}^2+m_{\rm e}^2} + k p_{\rm F})^2}
{(-k^2+\omega^2)^2 - 4 (\omega \sqrt{p_{\rm F}^2+m_{\rm e}^2} - k p_{\rm F})^2}
\right|.
\label{l3}
\end{eqnarray}
The quantity $L_1$ depends on the sign of
$D \equiv (\omega^2 - k^2)(k^2 - \omega^2 + 4 m_{\rm e}^2)$.
At $D \geq 0$ one has
\begin{eqnarray}
L_1 &=& \arctan \left[ \frac{- 2 m_{\rm e} k p_{\rm F}
+ (k^2 + 2 m_{\rm e} \omega - \omega^2)
\left(\sqrt{p_{\rm F}^2+m_{\rm e}^2} - m_{\rm e} \right)}
{p_{\rm F} \, \sqrt{(\omega^2-k^2) \, (k^2 - \omega^2 + 4 m_{\rm e}^2)}} \right]
\nonumber \\
&+& \arctan \left[ \frac{2 m_{\rm e} k p_{\rm F}
+ (k^2 + 2 m_{\rm e} \omega - \omega^2)
\left(\sqrt{p_{\rm F}^2+m_{\rm e}^2} - m_{\rm e} \right)}
{p_{\rm F} \, \sqrt{(\omega^2-k^2) \,
(k^2 - \omega^2 +4 m_{\rm e}^2)}} \right]
\nonumber \\
&+&\arctan \left[
\frac{- 2 m_{\rm e} k p_{\rm F} + (k^2 - 2 m_{\rm e} \omega - \omega^2)
\left(\sqrt{p_{\rm F}^2+m_{\rm e}^2} - m_{\rm e} \right)}
{ p_{\rm F} \, \sqrt{(\omega^2-k^2) \, (k^2 - \omega^2 + 4 m_{\rm e}^2)} }
\right]
\nonumber \\
&+&\arctan \left[ \frac{2 m_{\rm e} k p_{\rm F}
+ (k^2 - 2 m_{\rm e} \omega - \omega^2)
\left(\sqrt{p_{\rm F}^2+m_{\rm e}^2} - m_{\rm e} \right)}
{p_{\rm F} \,\sqrt{(\omega^2-k^2) \, (k^2 - \omega^2 + 4 m_{\rm e}^2)}} \right].
\label{l1_1}
\end{eqnarray}
At $D<0$
\begin{equation}
L_1 = \frac{i}{2} \, \ln \left|
\frac{\left[(k^2-\omega^2)\sqrt{p_{\rm F}^2 + m_{\rm e}^2} +
p_{\rm F} \sqrt{(k^2-\omega^2)
(k^2-\omega^2+4 m_{\rm e}^2)}\right]^2 - 4 m_{\rm e}^4 \omega^2}
{\left[(k^2-\omega^2)\sqrt{p_{\rm F}^2 + m_{\rm e}^2} -
p_{\rm F} \sqrt{(k^2-\omega^2)
(k^2-\omega^2+4 m_{\rm e}^2)}\right]^2 - 4 m_{\rm e}^4 \omega^2} \right|.
\label{l1_2}
\end{equation}
Note, that equations (\ref{epsldeg}) and (\ref{epsdegtr})
for the dielectric functions agree with the well known results
of Jancovici (1962) only at $D<0$
(see his equations A1 and A4).
At $D \geq 0$ his expressions (A1) and (A4) are formally
inapplicable (the real part of the dielectric functions
in these equations becomes complex).
In this case one should use our equations
(\ref{epsldeg}) and (\ref{epsdegtr}).

In addition, it may be useful to note,
that the Jancovici's definition
of the transverse dielectric function differs
from a generally accepted one.
His dielectric function $\varepsilon_{\rm t}^{\rm Janc}$
is related to our dielectric function by
$\varepsilon_{\rm t}^{\rm Janc}
=(k^2-\omega^2 \varepsilon_{\rm t})/(k^2-\omega^2)$.

\section[]{Description of a table of
our numerical results}
\label{appendixB}
The results of our numerical calculations
are summarized in the table
(file table.dat) which can be found
on the web:
http://www.ioffe.ru/astro/NSG/plasmon/table.dat.

The table consists of seven columns.
In the first column, we present
${\rm lg}T$ (in kelvins);
in the second column we give
${\rm lg}({\tilde \rho})=\rho/\mu_{\rm e}$ (g cm$^{-3}$);
in the third and fourth columns we present, respectively,
the emissivities $Q_{\rm t}$ and $Q_{\rm l}$
(erg s$^{-1}$ cm$^{-3}$) due to decay
of transverse and longitudinal plasmons;
the fifth column is the plasma frequency
$\omega_{\rm p}$ (s$^{-1}$),
which is numerically calculated
from the exact dispersion relations (\ref{disprel})
[not using the Braaten-Segel approximation];
the sixth column is the same plasma frequency
but calculated from equation (\ref{wp})
[the Braaten-Segel approximation].
Finally, in the seventh column we present
the characteristic dimensionless velocity of electrons
$v_{*}=\omega_1/\omega_{\rm p}$ in units of $c$,
calculated in the Braaten-Segel approximation
(i.e. by making use of equations \ref{wp} and \ref{w1}
for $\omega_{\rm p}$ and $\omega_1$, respectively).

\bsp

\label{lastpage}

\end{document}